\documentclass[12pt]{article}
\usepackage{amsmath}
\usepackage{latexsym}
\usepackage{amssymb}
\usepackage{a4wide}
\usepackage{bbm}
\usepackage{epsfig}
\usepackage{axodraw,graphicx,graphics,here,subfigure,rotating}
\usepackage{footnote}
\usepackage{floatflt} 
\usepackage{color}
\newcommand{\tr}{\text{tr}}

\newcommand{\re}[1]{~(\ref{#1})}
\newcommand{\Cas}{C_2(\Nc)}

\renewcommand{\bar}{\overline}
\newcommand{\pt}{\ensuremath{\partial_{t}}}

\newcommand{\Nc}{\ensuremath{\textrm{N}_{\textrm{c}}}}

\newcommand{\fss}[1]{#1\!\!\!/}

\newcommand{\I}{\text{i}}

\newcommand{\case}[2]{{\scriptstyle \frac{#1}{#2}}}
\newcommand{\Gk}{\Gamma_k}

\newcommand{\yb}{\bar{\psi}}
\newcommand{\Zy}{Z_{\psi}}
\newcommand{\ZF}{Z_{\text{F}}}
\newcommand{\ZB}{Z_{\text{B}}}
\newcommand{\pat}{\partial_t}

\newcommand{\SP}{\,(\text{S--P})}

\newcommand{\VAp}{\,(\text{V+A})}
\newcommand{\VAm}{\,(\text{V--A})}
\newcommand{\VAad}{\,(\text{V--A})^{\text{adj}}}
\newcommand{\VAn}{[2\!\VAad\!+({1}/{\Nc})\!\VAm]}

\newcommand{\etaF}{\eta_{\text{F}}}
\newcommand{\etaB}{\eta_{\text{B}}}
\newcommand{\ebar}{\bar{e}}
\newcommand{\gbar}{\bar{g}}
\newcommand{\Nf}{\ensuremath{\textrm{N}_{\text{f}}}}
\newcommand{\lambdah}{\hat{\lambda}}
\newcommand{\lp}{\hat{\lambda}_{+}}
\newcommand{\lm}{\hat{\lambda}_{-}}
\newcommand{\lsc}{\hat{\lambda}_{\sigma}^{\text{c}}}
\newcommand{\lsf}{\hat{\lambda}_{\sigma}}
\newcommand{\lva}{\hat{\lambda}_{\text{VA}}}

\newcommand{\blp}{\bar{\lambda}_{+}}
\newcommand{\blm}{\bar{\lambda}_{-}}

\newcommand{\blsf}{\bar{\lambda}_{\sigma}}
\newcommand{\blva}{\bar{\lambda}_{\text{VA}}}

\newcommand{\fsl}[1]{#1\!\!\!\!/}
\newcommand{\lF}{l_1^{\text{(F)},4}}
\newcommand{\lFB}{l^{\textrm{(FB)},4}_{1,2}}
\newcommand{\lFBo}{l^{\textrm{(FB)},4}_{1,1}}

\newcommand{\lsh}{\hat{\lambda}_{\sigma}}
\newcommand{\LUV}{\Lambda_{\text{UV}}}

\begin{document}

$\text{}$

\vspace{-2.3cm}

{\hfill HD-THEP-03-60}
 
\vspace{1.5cm}

\centerline{\Large\bf Towards a renormalizable standard model}

\vspace{2mm}

\centerline{\Large\bf without fundamental Higgs scalar}

\vspace{.8cm}



\centerline{Holger Gies${}^1$, Joerg Jaeckel${}^2$ and Christof
Wetterich${}^3$}

\vspace{.6cm}

\centerline{\small\it Institut f\"ur theoretische Physik,
  Universit\"at Heidelberg,}
\centerline{\small\it Philosophenweg 16, D-69120 Heidelberg,
  Germany}
\centerline{\small\it \quad ${}^1$ E-mail:
  h.gies@thphys.uni-heidelberg.de}
\centerline{\small\it \quad ${}^2$ E-mail:
  j.jaeckel@thphys.uni-heidelberg.de}
\centerline{\small\it \quad ${}^3$ E-mail:
  c.wetterich@thphys.uni-heidelberg.de}

\begin{abstract}
\noindent We investigate the possibility of constructing a
renormalizable standard model with purely fermionic matter
content. The Higgs scalar is replaced by point-like fermionic
self-interactions with couplings growing large at the Fermi scale.
An analysis of the UV behavior in the point-like approximation
reveals a variety of non-Gau\ss ian fixed points for the fermion
couplings. If real, such fixed points would imply nonperturbative
renormalizability and evade triviality of the Higgs sector. For
point-like fermionic self-interactions and weak gauge couplings,
one encounters a hierarchy problem similar to the one for a
fundamental Higgs scalar.
\end{abstract}

\section{Introduction}
The standard model of elementary particle physics is remarkably
successful in describing physics up to a scale of the order of several
hundred GeV. Still it faces a number of shortcomings, such as the
abundance of parameters and their origin, which become particularly
prominent in flavor physics or neutrino physics. In addition, there
are two problems that appear to point more strongly to the fact that
the standard model cannot be a fundamental theory valid for
arbitrarily short distance scales: first, the problem of triviality of
the scalar and abelian gauge sector, and second, the gauge hierarchy
problem.

Let us start with the hierarchy problem. If one assumes that the
standard model is valid up to some high scale $\LUV$ (e.g., a scale of
grand unification, $\textrm{M}_{\textrm{GUT}}\sim
10^{16}\textrm{GeV}$, or even the Planck scale), one is immediately
confronted with two immensely different scales in the theory -- the
electroweak symmetry-breaking scale $\textrm{M}_{\textrm{EW}}\sim
100\textrm{GeV}$ and $\LUV$. A realization of this enormous hierarchy
in the context of the standard model requires a highly exceptional
choice of parameters (`` fine-tuning''). This can be seen from the
quadratic renormalization of the scalar mass term in the Higgs sector
which naively receives corrections of the order $\LUV^{2}\gg
m^{2}_{\textrm{Higgs}}$ when quantum fluctuations between $\LUV$ and
$\textrm{M}_{\textrm{EW}}$ are integrated out. Thus
$m_{\textrm{Higgs}}$ at the UV scale must be extremely fine-tuned in
order to cancel most of the quantum corrections and produce a Higgs
which is much lighter than $\LUV$. It should be stressed that the
hierarchy problem is not a problem of principle, but rather a problem
of likeliness. Furthermore, it is sufficient that a suitable
renormalized parameter at the scale $\LUV$ be close to a particular
critical value, i.e., (partial) fixed point. No fine-tuning ``order by
order'' is necessary in a suitable renormalization group treament
\cite{Wetterich:1983bi}. Also the existence of the (partial) fixed
point is related to an enhanced (approximate) dilatation symmetry of
the quantum system and therefore natural. Nevertheless, the necessity
of an explanation for a ``close-to-critical value'' at $\LUV$ remains.

Triviality of the scalar and the abelian sector is a fundamental
problem. If we require a theory to be valid up to arbitrarily high
momentum scales, we have to take the limit $\LUV\to\infty$. Trivial
quantum field theories become noninteracting in this limit, since
their renormalized couplings vanish. Viewed differently, if we start,
for instance, with an interacting Higgs sector at low energies, the
(perturbative) running of scalar coupling approaches a singularity
(Landau pole), signaling the possible break-down of the theory and the
onset of new physics. Triviality in the flow of a renormalized
coupling that is observed experimentally at a nonzero value
experimentally unequivocally means that the standard model cannot be
continued to arbitrarily short distances. This obvious shortcoming of
a ``fundamental standard model'' is often not considered as a major
problem in practice: unification with gravity suggests that the
standard model has to be extended at the Planck scale anyhow.

In many attempts at formulating theories beyond the standard model,
the hierarchy and triviality problem are not considered
simultaneously. For the hierarchy problem, new physical mechanisms are
invoked to already operate on a low scale, say $\mathcal{O}$(TeV), in
order to strongly modify the renormalization group (RG) behavior of
the system. Prominent examples of these scenarios are supersymmetry
\cite{Nilles:1983ge}, technicolor \cite{Farhi:1980xs} or the ``little
Higgs'' \cite{Arkani-Hamed:2001nc}. Common to all of these is the need
of additional physical parameters instead of their desired
reduction. As the main hope, no parameter needs to be fine-tuned at
the high scale $\LUV$ and the system is (marginally) RG stable towards
the infrared (IR). The triviality problem is finally circumvented by
embedding the standard model in a nontrivial theory (as the abelian
U(1) is embedded in a nonabelian gauge group of a GUT), or by
referring to a framework that goes far beyond quantum field theory,
such as string theory.

This letter represents an attempt at tackling these problems from a
different viewpoint. Instead of modeling new physics with additional
parameters at the TeV scale (IR viewpoint), we are looking for quantum
field theoretic systems defined at a high scale $\LUV$ that have the
potential to remain RG stable down to low scales (UV viewpoint). The
desired systems should not have more relevant parameters than the
standard model; in particular, no new parameters should be required
for the system to become unstable near the Fermi scale (as is the
case, e.g., with soft-breaking terms in SUSY theories).

The hierarchy problem is deeply rooted in the scalar nature of the
Higgs sector. In the present work we avoid the concept of fundamental
scalars and consider only fermionic matter fields. The Higgs scalar is
then interpreted as a fermionic bound state, whereas the Higgs
condensate arises from dynamical symmetry breaking accompanied by a
condensation of fermion bilinears. This behavior is well known from
the Nambu--Jona-Lasinio (NJL) model \cite{E} and systems alike, which
have already frequently been used for modeling the Higgs sector, as in
topquark-condensation scenarios \cite{A}. In fact, from a
model-building point of view, our systems closely resemble those with
topquark condensation. However, we focus on the UV behavior, which is
an inherently nonperturbative problem.

At first sight, it seems that fermionic self-interactions even worsen
the theoretical objections, since such couplings are not
perturbatively renormalizable in $D=4$ dimensional spacetime. This
means that a quantum theory cannot consistently be constructed by an
expansion around zero coupling (Gau\ss ian fixed point). Nevertheless,
perturbative renormalizability is not a necessary criterion for
constructing an interacting field theory. Also perturbatively
nonrenormalizable theories can be fundamental and mathematically
consistent down to arbitrarily small length scales, as proposed in
Weinberg's ``asymptotic safety'' scenario \cite{Weinberg:1976xy}. This
scenario assumes the existence of a non-Gau\ss ian (=nonzero) UV fixed
point under the renormalization group (RG) operation at which the
continuum limit can be taken. The theory is ``nonperturbatively
renormalizable'' in Wilson's sense. If the non-Gau\ss ian fixed point
is IR repulsive only for a finite number of renormalized couplings,
the RG trajectories along which the theory can flow for
$M_{\text{EW}}/\LUV\to 0$ are labeled by only a finite number of
physical parameters. Then the theory is as predictive as any
perturbatively renormalizable theory, and high-energy physics can be
well separated from low-energy physics without tuning a large number
of parameters. Finally, the triviality problem is absent by
construction.

The issue of the gauge hierarchy problem is related to the relevant
couplings and the associated anomalous dimensions. For definiteness,
let us consider one particular small deviation\footnote{More
precisely, $v$ is an eigenvector of the stability matrix. We take $v$
to be a dimensionless renormalized parameter. In the case of a mass
$M$, this corresponds to $M/k$.} $v$ from the fixed point which
depends on the renormalization scale $k$ according to a generalized
``anomalous dimensions'' $\Theta$,
\begin{equation}
\pat v=k\partial_k v=-\Theta\, v. \label{0}
\end{equation}
The solution $v\sim k^{-\Theta}$ implies for large positive $\Theta$
that $v$ must be tiny at $\LUV$ if it is of order one for
$k=M_{\text{EW}}$. This is the fine-tuning problem. For positive
$\Theta$ (not very close to zero) $v$ is called relevant parameter and
we conclude that a fine-tuning problem is connected to every relevant
parameter. (We note that for a perturbative expansion the exact
location of the fixed point may depend on the order of the
approximation. Also $\Theta$ may be expressed in a perturbative series
and the right-hand side may contain higher powers of $v$.  All this
does not affect the conclusion that precisely one small parameter
$v(\LUV)$ is needed (one fine-tuning) for every relevant coupling.) On
the other hand, if $\Theta$ is very close to zero or vanishes (in this
case $v$ may depend logarithmically on $k$), $v$ is called a marginal
coupling. (An example is the gauge coupling $g$ near a fixed point
with $g=0$.) No extreme fine-tuning is needed for marginal couplings,
since $v(M_{\text{EW}})$ and $v(\LUV)$ are of a similar order of
magnitude. In consequence, a model with only marginal couplings has no
gauge hierarchy problem.  Within the standard model, this type of
solution for the gauge hierarchy problem has been proposed in
\cite{Bornholdt:1992up}.  Known examples for the quantum field
theories with only marginal couplings (besides the irrelevant ones)
are four-dimensional pure non-abelian gauge theories near the Gau\ss
ian fixed point or two-dimensional scalar theories with global U(1)
symmetry near the fixed point corresponding to the Kosterlitz-Thouless
phase transition with essential scaling
\cite{Kosterlitz:xp,VonGersdorff:2000kp}. In statistical physics, a
fixed point with only marginal directions can be associated to
``self-organized criticality''.

In this work, we analyze a class of models with four-fermion
self-interactions with gauge and flavor symmetry. In order to quantize
the systems in a nonperturbative framework, we employ the exact
renormalization group formulated in terms of a flow equation for the
effective average action $\Gamma_k$ \cite{Wetterich:1993yh},
\begin{equation}
\pat\Gamma_k=\frac{1}{2}\, \text{STr}\, \pat R_k\,
(\Gamma_k^{(2)}+R_k)^{-1}, \quad t=\ln \frac{k}{\LUV}. \label{ERG}
\end{equation}
The latter is a free-energy functional that interpolates between the
bare action $\Gamma_{k=\LUV}= S$ and the full quantum effective action
$\Gamma_{k=0}$. Here, $R_k$ denotes a to some extent arbitrary
regulator function that specifies the details of the momentum-shell
integrations. With the flow equation\re{ERG}, it is possible to
analyze the space of action functionals and its fixed-point structure,
in order to look for quantizable and renormalizable theories. These
correspond to zeros of the right-hand side of Eq.\re{ERG} with a
suitable finite number of relevant parameters, characterizing the
small deviations from the fixed point. Equation\re{ERG} is a
functional differential equation, since the right-hand side contains
the second functional derivative $\Gamma_k^{(2)}$ (the full inverse
propagator). Approximate solutions can be found by suitable
truncations of the space of actions.

Within the approximation of point-like four-fermion interactions, we
indeed find a variety of non-Gau\ss ian fixed points that give rise to
new universality classes of interacting quantum field theories and
solve the triviality problem in the Higgs sector. Concerning the
hierarchy problem, however, we show that these fermionic theories are
at least as IR unstable as a scalar Higgs sector, so that a very
precise choice of initial conditions remains necessary (fine-tuning).

Upon the inclusion of gauge field dynamics, the picture does not
change qualitatively as long as the gauge couplings remain
perturbatively small. The influence of the gauge fields on the
fermionic sector is subdominant. Moreover, the fermionic
self-interactions do not modify the leading-order running of the gauge
couplings, as we show with the aid of modified Ward-Takahashi
identities. As a consequence, fermionic self-interactions cannot cure
the triviality problem of the abelian U(1) sector, e.g., by rendering
this gauge coupling asymptotically free.

While we have not found a solution to the gauge hierarchy problem so
far, we reveal instructive general aspects of the structure and
influence of fermionic self-interactions in models with flavor and
gauge symmetry. Our findings may be taken as a hint toward possible
directions in the search for a satisfactory renormalizable standard
model without a fundamental Higgs scalar. As a rather speculative
example, we demonstrate in the appendix how the possible existence of
a non-Gau\ss ian fixed point in the U(1) gauge coupling could strongly
influence and even stabilize the RG behavior of a fermionic Higgs
sector towards the infrared.

\section{Toy Modeling the Standard Model}

In the present work, we consider a theory of fermions with
self-interactions as well as gauge-field interactions. Let us start
with all possible interactions that are compatible with a
$\textrm{U}(1)\times\textrm{SU}(\Nc)$ gauge symmetry and a chiral
$\textrm{SU}(\Nf)_{\textrm{L}}\times \textrm{SU}(\Nf)_{\textrm{R}}$
flavor symmetry for $\Nf$ fermion species. The $\textrm{SU}(\Nc)$
simulates the non-abelian, asymptotically free part of the standard
model gauge group, while the $\textrm{U}(1)$ models the abelian part
with its triviality problem.  For simplicity, the gauge fields are
assumed to couple to left- and right-handed fermions in the same way
with the same charges. In comparison to the standard model, we have
neglected the electroweak $\textrm{SU}(2){}_{\text{L}}$ chiral gauge
interactions and the differences between the hyper charges under the
$\textrm{U}(1)$.  The reason for concentrating on the strong
$\textrm{SU}(\Nc)$ gauge group instead of the electroweak
$\textrm{SU}(2){}_{\text{L}}$ is the following: if an IR stabilized
system exists, we expect the strongest gauge interaction to be a good
candidate for a destabilizing influence near the Fermi scale. Even
though symmetry breaking first affects the electroweak sector, it may
be caused by fermionic self-interactions in combination with the
strong gauge sector. An inclusion of the remaining standard model
building blocks is, in principle, straightforward: one should add the
weak gauge interactions and consider additional possible four-fermion
interactions which are consistent with the reduced flavor symmetries
and the gauge symmetries.  The number of possible four-fermion
interactions increases very substantially, however, owing to the lack
of parity symmetry and reduced flavor symmetry. For $\Nc=3$, $\Nf=6$,
$\bar{e}=0$ our model corresponds to the standard model in the limit
of vanishing weak and hypercharge gauge couplings.

We concentrate on a simple truncation with point-like couplings and
include all possible four-fermion interactions obeying an
$\textrm{SU}(\Nc)\times \textrm{U}(1)$ gauge symmetry and an
$\textrm{SU}(\Nf)_{\textrm{L}}\times \textrm{SU}(\Nf)_{\textrm{R}}$
flavor symmetry\footnote{We note that only the four-fermion
interactions are manifestly invariant under local gauge
transformations for all possible choices of the couplings. Gauge
invariance of the remaining terms is governed by (modified)
Ward-Takahashi identities as discussed in Sect. \ref{gaugeint}.},
\begin{eqnarray}
\Gk\!\!\!&=&\!\!\!\!\int
\yb(\I\Zy\fss{\partial}+Z_{1}\bar{g}\fsl{A}+Z^{\textrm{B}}_{1}\bar{e}\fsl{B})
\psi+\frac{\ZF}{4}F^{\mu\nu}_z F_{\mu\nu}^z
    +\frac{\ZB}{4} B^{\mu\nu}B_{\mu\nu}
    +\frac{(\partial_{\mu}A^{\mu})^2}{2\alpha}
    +\frac{(\partial_{\mu}B^{\mu})^{2}}{2\alpha_{B}}
\label{equ::truncsym}\\
&&\!\!+\frac{1}{2} \Big[
  Z_{-}\blm\!\VAm + Z_{+}\blp \!\VAp
  +Z_\sigma\blsf\! \SP  +Z_{\textrm{VA}}\blva \VAn\! \Big]\!
\nonumber
.
\end{eqnarray}
Here $A_\mu=A^z T^z$, $F_{\mu\nu}=F_{\mu\nu}^z T^z$ denotes the
nonabelian gauge potential and field strength, and $B_\mu$,
$B_{\mu\nu}$ the abelian ones. The gauge-field kinetic terms are
accompanied by wave-function renormalizations $\ZF$ and $\ZB$, the
fermionic one by $\Zy$. Similarly, $Z_{1}$, $Z^{\textrm{B}}_{1}$, $Z_{+}$,
$Z_{-}$, $Z_{\textrm{f}}$ and $Z_{\textrm{VA}}$ are the vertex
renormalizations, whereas $\ebar$, $\gbar$, $\bar{\lambda}$ denote the bare
couplings. The renormalized (dimensionless) couplings are defined as
\begin{equation}
g=\frac{\bar{g}Z_{1}}{\ZB^{{1}/{2}}Z_{\psi}},
\quad e=\frac{\bar{e}Z^{\textrm{B}}_{1}}{\ZF^{{1}/{2}}Z_{\psi}},
\quad \hat{\lambda}=\frac{Z_{\lambda}k^{2}\bar{\lambda}}{\Zy^2}.
\end{equation}
We work in the Landau gauge, $\alpha=\alpha_{B}=0$, which is known to
be a fixed point of the renormalization group \cite{Ellwanger:1995qf}
and has the additional advantage that the fermionic wave function is
not renormalized in our truncation, such that we can choose $\Zy=1$.

The four-fermion interactions can be classified according to their
color and flavor structure. Color and flavor singlets are
\begin{eqnarray}
\VAm&=&(\yb\gamma_\mu\psi)^2 + (\yb\gamma_\mu\gamma_5\psi)^2,
\nonumber\\
\VAp&=&(\yb\gamma_\mu\psi)^2 - (\yb\gamma_\mu\gamma_5\psi)^2
\nonumber,
\end{eqnarray}
where color ($i,j,\dots$) and flavor ($a,b,\dots$) indices are
contracted pairwise, e.g., $(\yb\psi)\equiv (\yb_i^a \psi_i^a)$. The
operators of non-trivial color or flavor structure are denoted by
\begin{eqnarray}
\SP&=&(\yb^a\psi^b)^2-(\yb^a\gamma_5\psi^b)^2 \equiv
   (\yb_i^a\psi_i^b)^2-(\yb_i^a\gamma_5\psi_i^b)^2,\nonumber\\
\VAad&=&(\yb \gamma_\mu T^z\psi)^2 + (\yb\gamma_\mu\gamma_5 T^z\psi)^2,
\label{eq::colorflavor}
\end{eqnarray}
where we define $(\yb^a\psi^b)^2\equiv \yb^a\psi^b \yb^b \psi^a$,
etc., and $(T^z)_{ij}$ denotes the generators of the gauge group in
the fundamental representation. Owing to Fierz identities, the last
four-fermion structure in Eq.\re{equ::truncsym} can also be written
as
\begin{equation}
\VAn=(\yb_i\gamma_\mu\psi_j)^2 +(\yb_i\gamma_\mu\gamma_5\psi_j)^2
   \equiv (\yb^a\gamma_\mu\psi^b)^2 +(\yb^a\gamma_\mu\gamma_5\psi^b)^2.
\label{equ::nonsingl}
\end{equation}
The set of four different fermionic self-interactions occurring in
Eq.\re{equ::truncsym} forms a complete basis. Any other point-like
four-fermion interaction invariant under $\textrm{SU}(\Nc)\times
\textrm{U}(1)$ gauge symmetry and $\textrm{SU}(\Nf)_{\textrm{L}}\times
\textrm{SU}(\Nf)_{\textrm{R}}$ flavor symmetry can be decomposed into
these base elements by means of Fierz transformations.

Evaluating the RG flow equation in the limit of point-like
interactions and projecting the result onto our
truncation\re{equ::truncsym}, we obtain the following $\beta$
functions for the dimensionless couplings $\hat{\lambda}$:
\begin{eqnarray}
\pat\lm\!=\beta_-\!\!\!&=&\!\!\!2\lm
    -4v_4 \lFBo\left[ \left( \frac{3}{\Nc}g^2 -3 e^2 \right)\lm
            -3g^2 \lva \right]
    \label{eq:lm}\\
&&\!\!\!-\frac{1}{8}v_{4}\lFB\left[\frac{12+9\Nc^2}{\Nc^2}g^{4}
    +48 e^4 -\frac{48}{\Nc} e^2g^2 \right]
    \nonumber\\
&&\!\!\!-8 v_4\lF \Big\{-\Nf\Nc(\lm^2+\lp^2) + \lm^2-2(\Nc+\Nf)\lm\lva
       +\Nf\lp\lsf + 2\lva^2 \Big\},
        \nonumber\\
\pat\lp\!=\beta_+\!\!\!&=&\!\!\!2 \lp -4v_4\lFBo \left[ \left(
-\frac{3}{\Nc}g^2+3e^2\right)\lp\right]
    \label{eq:lp}\\
&&\!\!\!-\frac{1}{8}v_{4}\lFB\left[
    -\frac{12+3\Nc^2}{\Nc^2} g^{4}-48
    e^4+\frac{48}{\Nc} e^2 g^2 \right]
    \nonumber\\
&&\!\!\!-8 v_4 \lF \Big\{ - 3\lp^2 - 2\Nc\Nf\lm\lp
        - 2\lp(\lm+(\Nc+\Nf)\lva)
        \nonumber\\
        &&\quad\quad\quad\quad\quad\quad
    \quad\quad\quad\quad\quad\quad\quad\quad
        + \Nf\lm\lsf + \lva\lsf
        +\case{1}{4}\lsf{}^2 \Big\},
    \nonumber
\end{eqnarray}
\begin{eqnarray}
\pat\lsf\!=\beta_\sigma\!\!\!&=&\!\!\!
    2\lsf -4v_4 \lFBo \left[(6\Cas\, g^2+3e^2)\lsf
    -6g^2\lp \right]\label{eq:lsf}\\
&&\!\!\!-\frac{1}{4} v_4 \lFB \Big[ -\frac{24
    -9\Nc^2}{\Nc}\, g^4 + 48 e^2 g^2 \Big] \nonumber\\
&&\!\!\! -8 v_4 \lF
      \Big\{ 2\Nc \lsf{}^2  - 2\lm\lsf - 2\Nf\lsf\lva
        - 6\lp\lsf \Big\},
        \nonumber\\
\pat\lva\!=\beta_{\text{VA}}\!\!\!&=&\!\!\!
    2 \lva-4v_4 \lFBo \left[ \left( \frac{3}{\Nc}g^2-3e^2\right)\lva -3g^2\lm \right]
    \label{eq:lva} \\
&&\!\!\!-\frac{1}{8} v_4 \lFB \left[ -\frac{24 - 3\Nc^2}{\Nc} g^4 +48 e^2 g^2 \right]
    \nonumber\\
&&\!\!\! -8 v_4 \lF
  \Big\{ - (\Nc+\Nf)\lva^2 + 4\lm\lva
        - \case{1}{4} \Nf \lsf{}^2\Big\}.\nonumber
\end{eqnarray}
Here $\Cas=(\Nc^2-1)/(2\Nc)$ is a Casimir operator of the gauge group,
and $v_4=1/(32\pi^2)$. The $l$ quantities are positive constant
numbers of $\mathcal{O}(1)$ that characterize the regulator dependence
\cite{Berges:2000ew}. For better readability, we have written all
gauge-coupling-dependent terms in square brackets, whereas fermionic
self-interactions are grouped inside curly brackets. For small gauge
couplings, the running of $g$ and $e$ is governed by their standard
perturbative $\beta$-functions -- this will be discussed in more
Detail in sect. \ref{gaugeint}.

\section{Fixed points for purely fermionic models}

\label{sec::vanishing}
A fixed point corresponds to a simultaneous zero of all
$\beta$-functions.  Each fixed point defines a (nonperturbatively)
renormalizable theory within our truncation. Each fixed point
furthermore constitutes its own universality class.  Let us first
analyze the RG flow of the fermionic couplings $\hat\lambda_i$ given
above in the simplified context of vanishing gauge couplings,
$g^2,e^2\to 0$. Then, in the point-like approximation the $\beta$
functions are all of the same form:
\begin{equation}
\pat \hat\lambda_i = \beta_i(\hat\lambda)
=(d-2)\lambda_i + \lambdah_k A^{kl}_i \lambdah_l,
\label{eq:lstruc}
\end{equation}
where $A^{kl}_i$ are constant matrices which are symmetric in the
upper indices, and we generalize the right-hand side formally to $d$
dimensional spacetime. For fixed $\lambdah_{j\neq i}$, the $\beta$ function
for $\lambdah_i$ corresponds graphically to a parabola, such that the
fixed-point equation $\pt\hat{\lambda}_i=\beta_i(\hat\lambda_{j\neq
i};\hat\lambda_i) =0$ has exactly two (possibly complex or degenerate)
solutions for $\hat\lambda_i$. Since our truncation
Eq.\re{equ::truncsym} has 4 fermionic couplings, we expect up to
$2^4=16$ different fixed points. A computer-algebraical inspection of
Eq.\re{eq:lstruc} indeed reveals these 16 fixed points, all of which are
real and therefore physically acceptable\footnote{This is a
particularity of the $\textrm{SU}(\Nf)_{\textrm{L}}\times
\textrm{SU}(\Nf)_{\textrm{R}}$ flavor symmetry. For instance, a less
restrictive SU$(\Nf)_{\text{V}}$ flavor symmetry does allow for 6
different four-fermion couplings, implying $2^6=64$ fixed points, only
44 of which are real and physically acceptable.}; we do not find
degeneracies.

Consequently, in this framework each fixed point serves as a
possibility to define a fundamental renormalizable quantum system in
which the continuum limit can be taken. The triviality problem is thus
absent; however, the hierarchy problem remains in these purely
fermionic systems in the point-like limit. This can be shown by
studying the stability matrix $B_{i}{}^{j}$, defined by the
derivatives of the $\beta$ functions at the fixed point
\begin{equation}
B_{i}{}^{j}=\frac{\partial
    \beta_i}{\partial\lambdah_{j}}\Bigg|_{\hat\lambda_\ast}
    =(d-2)\delta_{i}^{j}+2\lambdah_{\ast k}A_{i}^{kj}.
\end{equation}
The eigenvalues of the stability matrix and the associated
eigenvectors $v$ govern the evolution of small deviations from the
fixed point according to $B v=-\Theta v$. (We denote by $\Theta$ the
negative of the eigenvalues). In turn, this determines the running of
the couplings in the fixed-point regime,
$(\hat\lambda-\hat\lambda_\ast)\sim (\LUV/k)^\Theta$. Therefore, large
positive $\Theta$ implies a rapid growth of the couplings towards the
infrared and corresponds to a strongly relevant RG direction,
indicating IR instability such as a scalar mass term.

In the present problem, a large positive eigenvalue $\Theta=(d-2)$ of
the stability matrix exists for each fixed point
$\hat\lambda_\ast$. This can been seen from the following argument:
let $\lambdah_{\ast}$ be a solution of the fixed point equation,
\begin{equation}
\label{equ::fixed}
\pt\lambdah_{\ast i}
   =(d-2)\lambdah_{\ast i}
    +\lambdah_{\ast k}A_{i}^{kl}\lambdah_{\ast l}=0
\quad, \forall i.
\end{equation}
Acting with $B_{i}{}^{j}$ on $\lambdah_{\ast j}\neq 0$ we have
\begin{eqnarray}
B_{i}{}^{j}\,\lambdah_{\ast j}
   &=&(d-2)\lambdah_{\ast i}
    +2\lambdah_{\ast j}A_{i}^{jk}\lambdah_{\ast k}
    \nonumber\\
&=&-(d-2)\lambdah_{\ast i}+2((d-2)\lambdah_{\ast i}
    +\lambdah_{\ast j}A_{i}^{jk}\lambdah_{\ast k})
\nonumber\\
&=&-(d-2)\lambdah_{\ast i}, \label{eq:argument}
\end{eqnarray}
where we have used the fixed-point equation in the last step.  This
shows that $\lambdah_{\ast}$ itself is an eigenvector of the stability
matrix with the eigenvalue $-(d-2)$, hence $\Theta=(d-2)$. For $d=4$
we have therefore at least one ``quadratically renormalizing''
relevant direction in the fixed-point regime, which is the same RG
behavior as for a system with a fundamental Higgs scalar. A very
precise choice for the initial conditions at the high scale (GUT
scale) is required in order to separate the high scale from the
symmetry-breaking scale (Fermi scale).

The second lesson to be learned from Eq.\re{eq:argument} is that there
cannot be a purely infrared attractive fixed point besides
$\lambdah=0$ in this truncation. In other words, all remaining 15
fixed points can be used for defining an interacting continuum
limit, which requires at least one IR repulsive (or marginal)
direction.

\begin{figure}[t]
\begin{center}
\scalebox{0.55}[0.55]{
\begin{picture}(300,230)(0,20)
\SetOffset(4.5,0)
\SetColor{Red}{
\Vertex(158,60.5){4}
\Vertex(158,235){4}
\Vertex(77,161){4}
\Vertex(243,149){4}}
\Text(-5,250)[]{\scalebox{1.6}[1.6]{$\lp$}}
\Text(255,-10)[]{\scalebox{1.6}[1.6]{$\lm$}}
\includegraphics{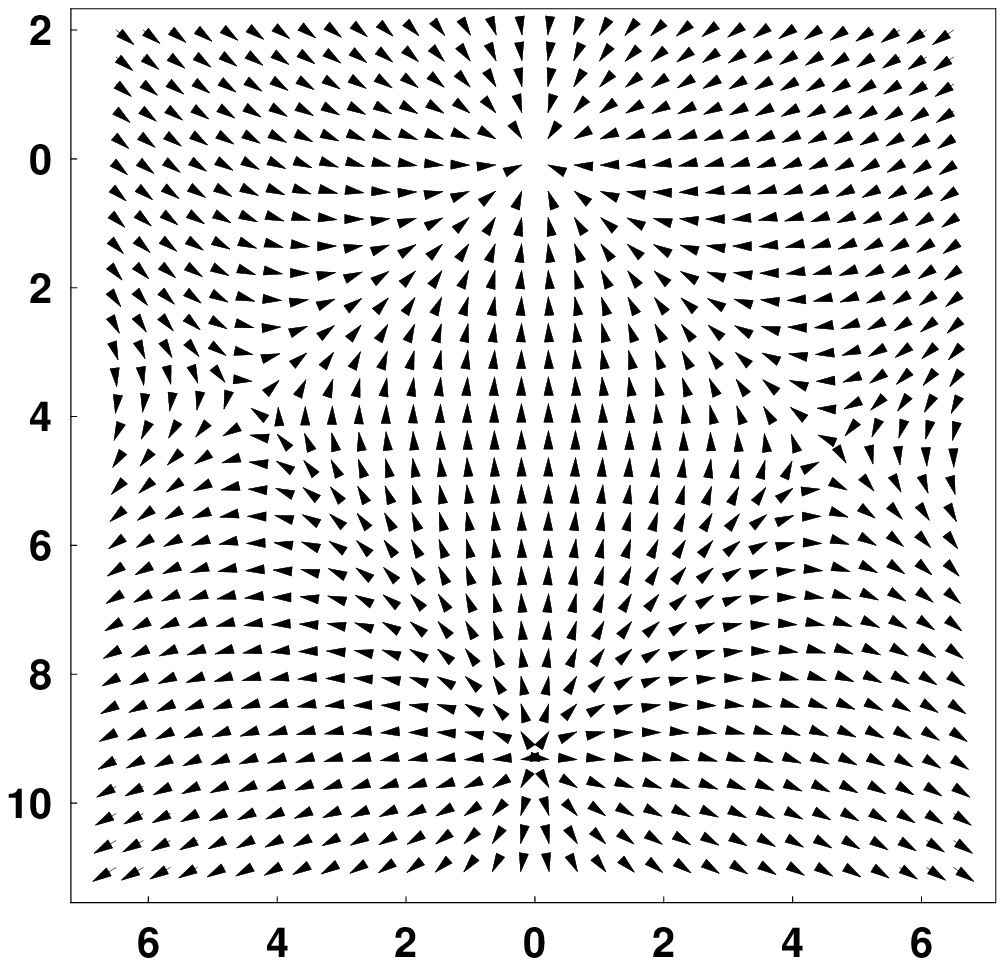}
\end{picture}}
\quad\qquad
\scalebox{0.65}[0.65]{
\begin{picture}(300,230)(0,20) 
\includegraphics{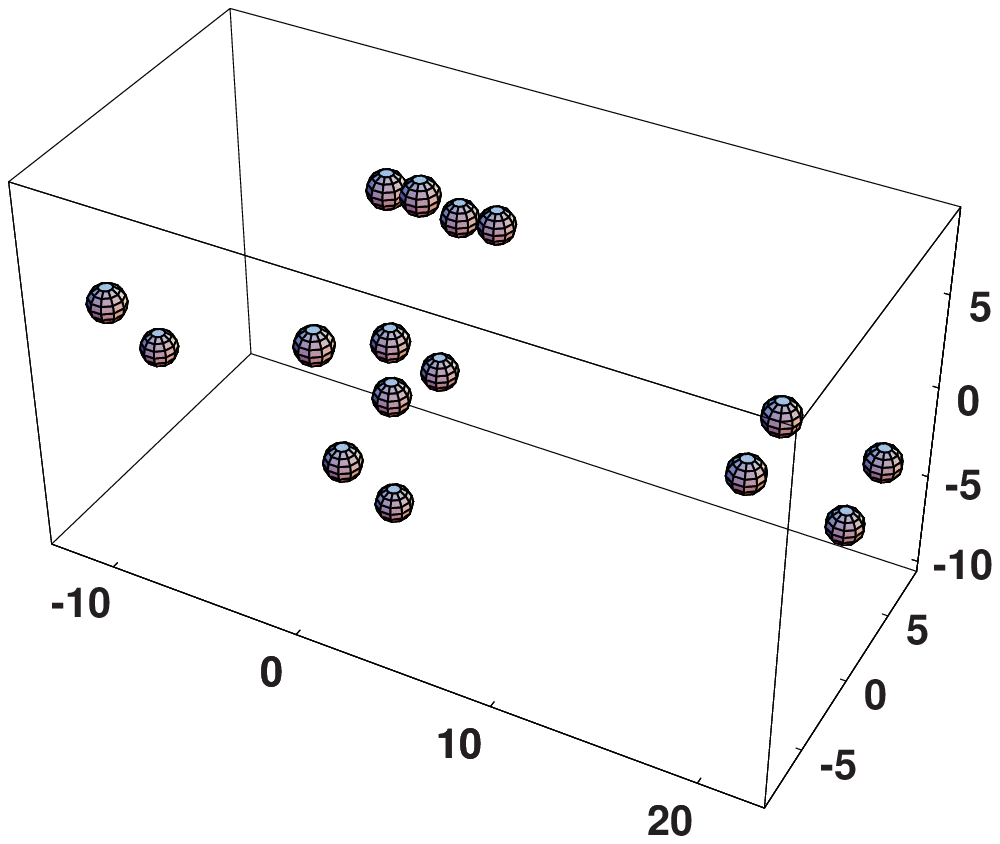}
\Text(-8,30)[]{\scalebox{1.6}[1.6]{$\lp$}}
\Text(15,130)[]{\scalebox{1.6}[1.6]{$\lm$}}
\Text(-200,20)[]{\scalebox{1.6}[1.6]{$\lsf$}}
\end{picture}}
\end{center}
\caption{Left panel: RG flow in the ($\lp$,$\lm$) subsystem. The
filled circles (red) mark the fixed points, and the arrows show the
direction of the flow toward the infrared. It is easy to see that only
the Gau\ss ian fixed point, $\lp=\lm=0$, is IR stable.  All others
have at least one relevant, IR repulsive direction. Right Panel: all
16 fixed points of the full system projected onto the $(\lp,\lm,\lsf)$
subspace.}
\label{2dim}
\end{figure}

The fixed points can be classified further according to their number
of relevant and irrelevant directions, constituting the number of
physical parameters. In our case, the number of fixed points with $j$
relevant directions turns out to be equal to the binomial coefficient
$\left(
\begin{array}{c} 4 \\ j \end{array} \right)$. There is one IR
stable fixed point (only irrelevant directions) which is the Gau\ss
ian fixed point, and also exactly one fixed point with only relevant
directions which is IR repulsive in all directions in this
truncation. All other fixed points have relevant and irrelevant
directions. This is illustrated in the left panel of Fig. \ref{2dim}
using the subsystem $\lp$, $\lm$ as a simple example (further
parameters: $d=4$, $\Nc=3$, $\Nf=6$, and linear cutoff functions
\cite{Litim:2001up} ($\lF=1/2,\lFBo=1,\lFB=3/2$)). The right panel of
Fig. \ref{2dim} displays all 16 fixed points projected onto the
$(\lp,\lm,\lsf)$ subspace.

\section{Gauge interactions}
\label{gaugeint}

\begin{figure}[t]
\begin{center}
\scalebox{0.65}[0.65]{
\fbox{
\begin{picture}(160,80)
\SetOffset(-87,-20)
\ArrowLine(130,60)(95,95)
\ArrowLine(95,25)(130,60)
\Vertex(130,60){2}
\ArrowArc(160,60)(-30,0,180)
\ArrowArc(160,60)(-30,180,0)
\Vertex(190,60){2}
\Photon(190,60)(235,60){5}{4}
\end{picture}}}
\end{center}
\caption{Correction to the gauge-boson--fermion vertex (fermions solid
with arrow, gauge boson wiggled) in the presence of a four-fermion
interaction.}
\label{fig::gauge}
\end{figure}
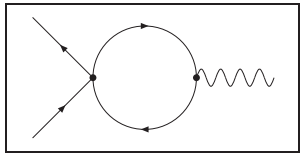

Let us now include the gauge interactions in our considerations. For
this, we need the running of the gauge couplings which we derive from
the fermion--gauge-field vertex $\Gamma_\mu$. For instance, in the
abelian case, the general form of this vertex is
\begin{equation}
\Gamma_{\yb \psi B}=\ebar\int_{q_1,q_2} \yb(q_2)\,
\Gamma_\mu(q_2,q_1)\, B_\mu(q_2-q_1)\, \psi(q_1), \label{vertex}
\end{equation}
from which we define the renormalized coupling $e$ in the Thompson limit,
\begin{equation}
\lim_{p\to0} \Gamma_\mu(q,q+p)=\Zy\,\frac{e}{\ebar}\, \ZF^{1/2}\,
\gamma_\mu=Z^{\textrm{B}}_{1}\gamma_{\mu}, \label{runcoup}
\end{equation}
and similarly for the nonabelian coupling $g$.  (Here we included the
fermion wave-function renormalization $\Zy$ for full generality, but,
as we already mentioned, with Landau gauge we have $\Zy=1$ in our
truncation.)  Within our truncation, the flow equation for the vertex
results in the $\beta$ functions
\begin{eqnarray}
\pat g^2 &=& \etaF\, g^2
    - 8 v_4 \lF \Big[ \lsh-2\lm+\Nf \lsf -2\Nf \lva\Big] g^2,
    \label{eq:gq}\\\label{equ::oldgauge}
\pat e^2 &=& \etaB\, e^2
    - 8 v_4 \lF \Big[ \lsh-2\lm -2\Nf\Nc(\lp+\lm) \nonumber\\
&&\qquad\qquad\qquad\qquad\qquad
       +(\Nc\lsc+\Nf\lsf) -2(\Nc+\Nf)\lva \Big] e^2,
    \label{eq:eq}
\end{eqnarray}
where the standard one-loop coefficients are contained in the
anomalous dimensions of the gauge field,
\begin{eqnarray}
\etaF=-\frac{1}{\ZF} \pat \ZF=-4 v_4 b_0^{g^2}\, g^2, &&\quad
    b_0^{g^2}= \frac{11}{3}\, \Nc -\frac{2}{3} \, \Nf, \nonumber\\
\etaB=-\frac{1}{\ZB} \pat \ZB=-4 v_4 b_0^{e^2}\, e^2, &&\quad
    b_0^{e^2}= -\frac{4}{3}\, \Nf\Nc. \label{eq:oneloopcoef}
\end{eqnarray}
The additional $\hat{\lambda}$-dependent terms in Eqs.\re{eq:gq},\re{eq:eq}
arise from diagrams of the form shown in Fig.~\ref{fig::gauge}.
At first sight, it seems that these additional terms offer a new and
rich structure for the possible UV behavior of the system. For
instance, for a given non-Gau\ss ian $\hat{\lambda}$ fixed point
$\hat{\lambda}_\ast$, these terms are dominant in the small gauge
coupling limit. If the factor in square brackets in these terms is
positive for a given $\hat{\lambda}_\ast$, the corresponding gauge
coupling seems to be asymptotically free. This would offer a solution
to the triviality problem of the U(1) sector. Moreover, the interplay
of both terms on the right-hand side of these $\beta$ functions can
produce non-Gau\ss ian fixed points in the gauge couplings. This would
not only be a possible solution of triviality, but could also lead to
a sizeable reduction of the critical exponents and the hierarchy
problem by circumventing the argument of Eq.\re{eq:argument}.

However, these hopes can unfortunately not be confirmed, as quantum
field theory tells us in the following interesting
way. Equations\re{eq:gq} and\re{eq:eq} are not the only source of
information about the vertex that we can obtain from the flow-equation
formalism. The requirement of gauge invariance is encoded in a
constraint for the effective action: the Ward-Takahashi identity
(WTI). Since the regulator of the flow equation formalism is not
manifestly gauge invariant, it also contributes to the constraint,
leading to a modified Ward-Takahashi identity (mWTI)
\cite{Ellwanger:iz}.

For simplicity, let us analyze the mWTI arising from abelian
gauge symmetry. Employing the generator $\mathcal{G}$ of an
infinitesimal gauge transformation (in momentum space),
\begin{equation}
\mathcal{G}(p)=i p_\mu \frac{\delta}{\delta B_\mu(-p)} - i \ebar
  \left[  \int_q \psi(q) \frac{\delta}{\delta \psi(q-p)}
     -\int_q \yb(q) \frac{\delta}{\delta \yb(q+p)} \right],
\label{gaugegen}
\end{equation}
the mWTI can be written as
\begin{equation}
\mathcal{G}(p)\, \Gamma_k -\frac{i}{\alpha_B} \, p^2 p_\mu B_\mu(p)
=   -i \ebar\, \tr \int_q \big[R_k^\psi(q+p) \, G_{\psi\yb} (q+p,q)
-R_k^{\psi\,T}(q+p)\, G_{\yb^T\psi^T}(q+p,q) \big],\label{abelmWTI}
\end{equation}
where, e.g., $G_{\psi\yb}=(\Gamma^{(2)}+R_k)_{\psi\yb}^{-1}$ denotes
the $(\psi\yb)$ component of the propagator, and the $T$ symbol
indicates transposition in Dirac space. For vanishing regulator
$R_k\to 0$, the right-hand side of the mWTI is zero and we rediscover
the standard WTI. However, since we are dealing with a perturbatively
nonrenormalizable theory, the presence of the regulator is essential
in order to define the quantum theory, i.e., the Schwinger
functional. Therefore, the right-hand side of the mWTI should not only
be viewed as a technical complication, but as important information
about the structure of the theory.

Owing to its similar structure, the mWTI can be evaluated with the
same technology as the flow equation; the right-hand side, e.g., is
again of one-loop form with an exact propagator in the
loop. Information about the running gauge coupling can be found by
projecting the mWTI onto the operator $\sim \yb\psi$, yielding
\begin{equation}
e=\ebar\, \ZB^{-1/2}\, \left( 1-2v_4\lF \sum c^e_i \hat{\lambda}_i
\right), \label{mWTIe}
\end{equation}
i.e.,
\begin{equation}
\label{zrelation}
\frac{Z^{\textrm{B}}_{1}}{\Zy}=\left( 1-2v_4\lF \sum c^e_i
\hat{\lambda}_i \right), 
\end{equation}
where we abbreviated the combination
\begin{equation}
 \sum c^e_i \hat{\lambda}_i:=\lsh-2\lm -2\Nf\Nc(\lp+\lm)
       +(\Nc\lsc+\Nf\lsf) -2(\Nc+\Nf)\lva, \label{sumabb}
\end{equation}
as it also occurs in Eq.\re{eq:eq}. In ordinary QED, the term $\sim
\hat\lambda$ is not present and we end up with the standard result
that the running of the coupling corresponds to the running of the
gauge-field wave-function renormalization\footnote{With
Eq. \eqref{zrelation}, this corresponds to the standard result
$Z^{\textrm{B}}_{1}=Z_\psi$. We remark that from Eq. \eqref{zrelation}
alone only the ratio of $\Zy$ and $Z^{\textrm{B}}$ is accessible (and
physically relevant). Still, in approximations it can make a
difference if $\Zy$ or $Z^{\textrm{B}}_{1}$ is running. Consistent
with our choice of the Landau gauge, we set $Z_{\psi}=1$.}, $e=\ebar
\ZB^{-1/2}$. The presence of the fermionic interactions replaces this
simple relation by Eq.\re{mWTIe}.

By differentiation, we obtain the $\beta$ function
\begin{equation}
\pat e^2
{=}\etaB\, e^2
   -4 v_4 \lF\, \frac{e^2}{1-2v_4\lF \sum c^e_i \hat{\lambda}_i}
       \, \pat \sum c^e_i \hat{\lambda}_i. \label{betaeq}
\end{equation}
Here $\pat \sum c^e_i \hat{\lambda}_i$ means that we have to insert
the complete $\beta$ functions for the $\hat{\lambda}_i$ into
Eq.\re{betaeq}. Since $\pat \sum c^e_i \hat{\lambda}_i=2\sum c^e_i
\hat{\lambda}_i+\dots$ in a small-coupling expansion, we rediscover
the result of the flow equation \eqref{equ::oldgauge} in this
limit. In other words, flow equation and mWTI agree within the order
of our truncation, as they should. The mWTI, however, contains
considerably more information, since the non-monomial appearance of
the $\hat\lambda$'s suggests that the mWTI represents a resummation of
a larger class of diagrams.

This has important consequences for the RG behavior of the gauge
coupling, if compared to the possibilities that are offered by the
simpler form of Eq.\re{eq:eq}. At the non-Gau\ss ian $\hat\lambda$
fixed points, the $\hat\lambda$ flow vanishes, such that $\pat \sum
c^e_i \hat{\lambda}_i\to 0$. This implies that the running of the
gauge coupling in the vicinity of its Gau\ss ian fixed point is
determined by the standard one-loop term $\etaB e^2$ alone. The
fermionic self-interactions contribute only to higher
order.\footnote{A careful analysis reveals that the additional term in
Eq.\re{betaeq} is of the same order as the two-loop running of the
gauge coupling for small perturbations around the fixed point.} The
abelian gauge coupling can therefore not be rendered asymptotically
free by the influence of the four-fermion interactions. By a similar
argument, the additional term in Eq.\re{betaeq} does not facilitate
the existence of non-Gau\ss ian fixed points, $e_\ast\neq0$, in the
abelian gauge coupling within this truncation. Again, if the
$\hat\lambda$'s approach a fixed point, $\pat \sum c^e_i
\hat{\lambda}_i\to 0$ and we are left with the standard running only,
for which no non-Gau\ss ian fixed point is known.

To summarize, the four-fermion contributions to the running of the
abelian gauge coupling are not capable of solving the triviality
problem in our truncation.

Let us finally comment on the RG flow of the running SU($\Nc$)
coupling $g^2$. Although the mWTI for the nonabelian sector has a more
complex structure, the result for the four-fermion contribution to the
running gauge coupling has the same form as in Eq.\re{betaeq}. Near
the Gau\ss ian fixed point, the standard one-loop running holds and
the non-abelian gauge sector remains asymptotically free. Actually,
this finding is in line with another argument: we could equally
well define the running of the nonabelian gauge coupling by the
three-gauge-boson vertex. At one-loop order, there is no contribution
to the renormalization of this vertex from the four-fermion couplings
$\hat\lambda$. As a consequence, the usual one-loop $\beta$ function
governs the flow of this three-gluon vertex.


\section{Spontaneous symmetry breaking}

In order to illustrate the flow of the system from a non-Gau\ss ian
fixed point towards the regime of spontaneous symmetry breaking, we
select a fixed point with only one relevant direction (for instance,
the fixed point on the right-hand side or left-hand side in the left
panel of Fig.~\ref{2dim}). At the high scale $\LUV$, we specify
initial conditions such that all couplings in our truncation are
roughly in the vicinity of their fixed point values; no fine-tuning is
needed for this step. If we now compute the RG flow towards the
infrared, the system either relaxes towards the Gau\ss ian fixed point
with $\lambdah\to 0$ as long as the gauge couplings are weak; or the
system rapidly approaches the regime of spontaneous symmetry breaking
(SSB) which is signaled by diverging four-fermion couplings. In the
former case, the system is in the universality class of
SU$(\Nc)\times$U(1) gauge theory with $\Nf$ chiral fermions: in the
latter case we are dealing with a universality class characterized by
SSB. Finally, we fine-tune only one of the parameters of the initial
conditions such that the system is very close to the phase boundary on
the SSB side. This fine-tuning corresponds effectively to a
determination of the scale $k_{\text{SSB}}$ at which the system runs
into the SSB regime.

\begin{figure}[t]
\begin{center}
\begin{picture}(400,125) 
\put(170,-5){$t_{10}$}
\put(40,110){$\lambdah_i$}
\put(-5,0){\epsfig{figure=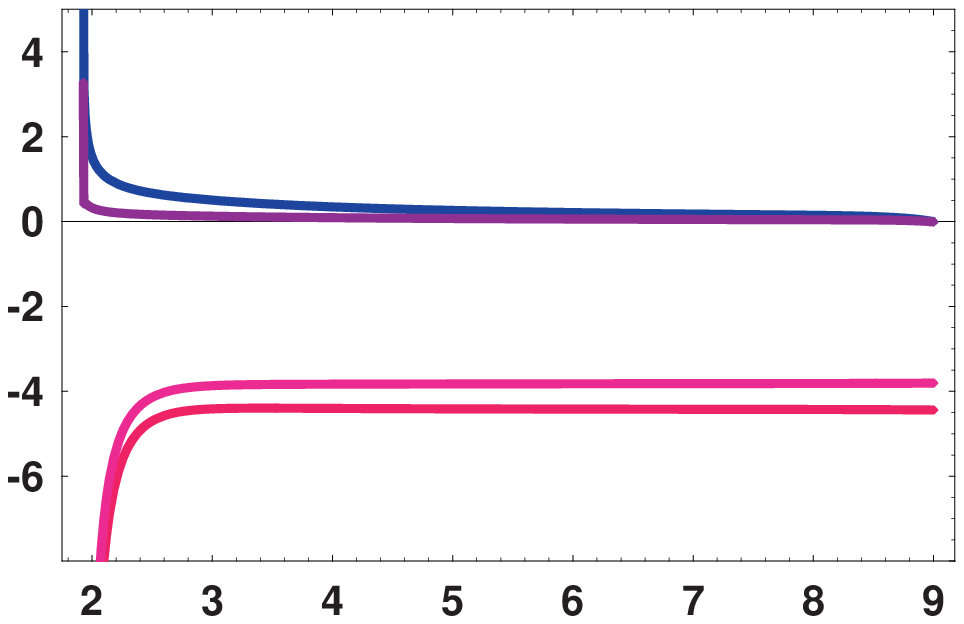,width=7cm}}
\put(373,-5){$t_{10}$}
\put(350,105){$g^2/(4\pi)$}
\put(350,52){$e^2/(4\pi)$}
\put(205,0){\epsfig{figure=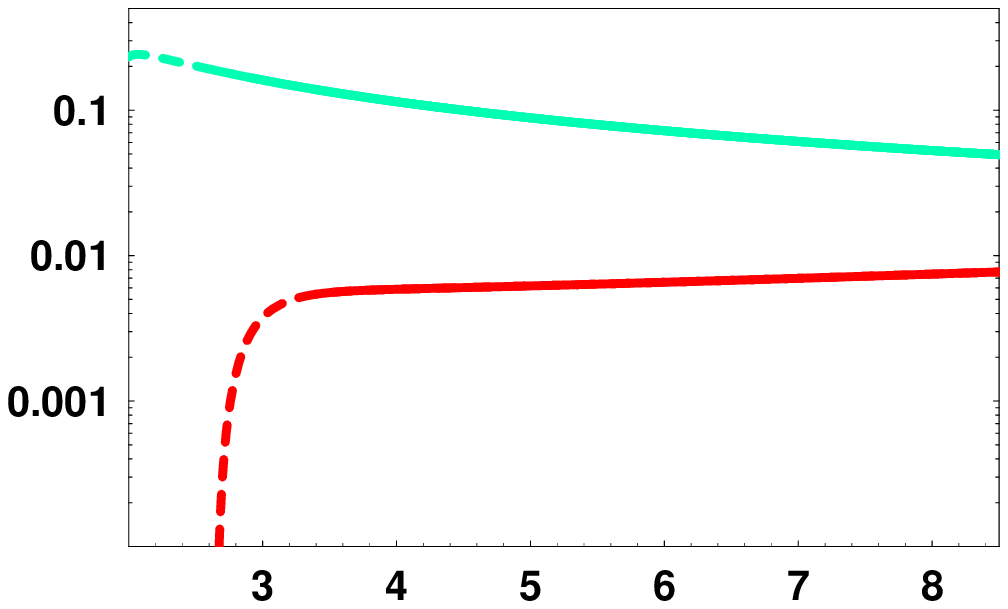,width=7cm}}
\end{picture}
\end{center}
\vspace{-.5cm}
\caption{Running couplings as a function of the renormalization scale
$k=10^{t_{10}}$ GeV.  Left panel: flow of the four-fermion couplings
$\lsf$, $\lva$, $\lm$, $\lp$ (from top to bottom); the divergence of
the couplings near the Fermi scale $10^{t_{10}}\sim 10^2$ signals the
approach to SSB. Right panel: flow of the gauge couplings; while the
$\lambdah_i$ are close to their fixed-point values, the gauge
couplings run according to standard perturbation theory. The rapid
behavior near the Fermi scale is likely to be an artefact of the
truncation (dashed lines).}
\label{figSSB}
\end{figure}

In Fig.~\ref{figSSB}, we display a particular solution to the flow
equation obtained in the afore-mentioned way for $\Nc=3$, $\Nf=6$ and
a linear regulator \cite{Litim:2001up}. We have adjusted the gauge
couplings roughly to their standard model values and fixed the
four-fermion couplings to the value of one of the four non-Gau\ss ian
fixed points with one relevant direction. Finally, we have fine-tuned
$\lva$ so that the SSB scale corresponds to the Fermi scale $\sim
10^2$ GeV. On the left panel, the running of the four-fermion
couplings from $k=10^9$ GeV down to the Fermi scale is depicted. Over
a wide range of scales, these couplings remain close to their
fixed-point values with a slight modulation induced by the logarithmic
increase of the gauge couplings. Near the Fermi scale, the running
induced by the relevant direction becomes fast and the four-fermion
couplings diverge, which signals the onset of SSB. With the present
truncation, however, we cannot enter the SSB regime where the dynamics
is governed by composite bosonic fluctuations on top of bosonic
condensates $\sim \langle \yb \psi \rangle$. A suitable description
can be given by means of partial bosonization under the RG flow
\cite{Gies:2001nw}. In the present case, this would relate, e.g., the
coupling $\lsf$ to a Yukawa coupling $h_\sigma$ and a mass term
$m_\sigma$ for the composite scalar, $\lsf\sim
h_\sigma^2/m_\sigma^2$. In this sense, the increase of
$\lambda_\sigma$ is associated with a decrease of the scalar mass
term, which eventually drops below zero and thus gives rise to SSB.

Using partial bosonization, one can moreover study the nature of the
condensate, whereas in the present purely fermionic description we
cannot distinguish the behavior of the various fermionic interaction
channels. Owing to the nonlinear interplay of the flow equations for
the couplings, all diverge simultaneously in this truncation. Even the
gauge couplings can be affected, as is the case in our example for the
abelian gauge coupling (see right panel of Fig.~\ref{figSSB}). This is
clearly an artefact of the present truncation, and we expect a
well-controllable flow once the threshold behavior is accounted for
by using the techniques of \cite{Gies:2001nw},\cite{Jaeckel:2003kf}.

\section{Conclusions}

In this letter, we have analyzed the RG behavior of a
standard-model-like system with purely fermionic matter content. The
Higgs sector is replaced by fermionic self-interactions which are
responsible for spontaneous symmetry breaking. Whereas the low-energy
side of our models is reminiscent, and in the spirit, of
topquark-condensation scenarios, we here concentrate on the UV
behavior of such systems, investigating their renormalizability and RG
stability in the framework of RG flow equations.

Within our truncation of point-like four-fermion interactions, we have
identified a large number of non-Gau\ss ian fixed points in the
fermionic interactions, each of which constitutes an independent
universality class with the given gauge and flavor symmetries at
hand. From the structure of the flow equations, we deduce that, for
$n$ independent four-fermion interactions, there exist up to $2^n$
fixed points. Each fixed point can serve to define an interacting
continuum limit. We find no sign of triviality in our truncation in
this fermionic Higgs sector. Contrary to the standard scalar Higgs
sector, the fermionic systems have the potential to be valid down to
arbitrarily small distances.

Furthermore, we have analyzed the RG stability of the model towards
the IR. As a result, all fixed points with non-vanishing four-fermion
interaction exhibit one RG relevant direction with critical exponent
2, i.e., renormalizing quadratically, similar to a fundamental
scalar. Therefore, our fermionic models suffer from the same hierarchy
problem as the conventional Higgs sector.

These findings are not modified by the inclusion of weakly coupled
gauge interactions which only induce small anomalous
dimensions. In turn, the fermionic self-interactions do not modify the
leading-order running of the gauge couplings, as we have shown with
the help of modified Ward-Takahashi identities. This inhibits a
solution of the triviality problem in the abelian gauge sector at the
Gau\ss ian fixed point.

In summary, we find that a fermionic Higgs sector has the potential to
be a truly renormalizable theory, removing the triviality problem of
a fundamental scalar Higgs. However, we have not been able to identify
a consistent resolution of the hierarchy problem within our
truncation. In a realistic scenario comprising the full standard-model
phenomenology, the number of physical parameters in our model would be
comparable to that of the standard model. The precise number will
depend on the particular choice of the non-Gau\ss ian $\lambda_\ast$
fixed point and its number of RG-relevant directions. Whether the
number of physical parameters can even drop below that of the standard
model then depends on the universality class associated with the
chosen fixed point. A determination of the physically acceptable
universality classes requires an analysis of their symmetry-breaking
properties. This is beyond the scope of the present truncation in
which all couplings diverge at the symmetry-breaking scale. These
low-energy properties can, however, easily be derived from an analysis
of the condensing bilinear fermion channels using partial bosonization
under the flow as described in \cite{Gies:2001nw},
\cite{Jaeckel:2003kf}.

At this point, let us comment on the stability of our results under a
change of the truncation. We have checked that higher fermionic
self-interactions do not modify our results in the point-like
limit. They neither remove the $\lambda_\ast$ fixed points nor
represent nonperturbatively renormalizable couplings themselves (the
latter would increase the number of physical parameters). Concerning
the gauge sector, we have studied a number of non-minimal
fermion--gauge-field couplings. None of them turns out to influence
the leading-order running of the gauge couplings in the weak-coupling
regime; the argument proceeds similarly to the one given in
Sect.~\ref{gaugeint} based on the mWTI.

The UV fixed points that we find for the fermionic couplings may be
viewed as a generalization of those UV fixed points that are known
from large-$\Nf$ studies of simple four-fermion interactions in
$d=2+1$ dimensions \cite{Rosenstein:pt}. In the latter case,
four-fermion interactions can be renormalized order by order in a
$1/\Nf$ expansion despite its seeming perturbative
nonrenormalizability. Our truncation exhibits these fixed points in
all dimensions $d>2$. However, we suspect that at least far beyond
four dimensions the fixed points may be an artefact of the truncation,
since here even the induced Yukawa couplings between fermions and
composite bosonic fluctuations become RG irrelevant by power-counting
arguments. In $d=4$, these Yukawa couplings are RG marginal by
power-counting: hence $d=4$ appears to be the critical
dimension. Large-$\Nf$ arguments are indeed in favor of logarithmic
triviality in $d=4$, a picture that receives some support from lattice
simulations with staggered fermions for a simple NJL model
\cite{Kim:pg}. However, the large-$\Nf$ approximation neglects the
anomalous dimension of the fermion which, even if tiny, can have a
large effect by changing marginal-irrelevant into marginal-relevant
operators. Since gauge interactions also contribute to the fermionic
anomalous dimension, purely fermionic lattice studies cannot be
conclusive for the models considered in the present work. Therefore,
$d=4$ lattice investigations with four-fermion as well as gauge
interactions would be desirable and may indeed be accessible with
recently developed algorithms \cite{Kim:2001am}.

In order to pursue this question further within the flow equation
framework, we have to go beyond the point-like limit. If the
qualitative picture developed so far in this simple truncation turns
out to be incomplete, we expect that strong modifications might arise
from the full momentum structure of the interaction. It may well be
that our non-Gau\ss ian $\lambda_\ast$ fixed points are only a
projection of a more general momentum-dependent interaction onto the
point-like limit. If so, it is natural to speculate that a strongly
momentum-dependent wave-function renormalization of the fermions could
even induce a large fermionic anomalous dimension. Then it would be
conceivable that the latter stabilizes the fermionic flow towards the
infrared. This would pave the way for a possible solution of the
hierarchy problem in models with purely fermionic matter
content. Concerning the hierarchy problem, another speculative
alternative based on the assumed existence of a non-Gau\ss ian fixed
point for the gauge couplings has been investigated in the
appendix. Whether or not one of these scenarios can indeed be realized
is subject to further nonperturbative studies for which RG flow
equations offer an appropriate framework.

\section*{Acknowledgment}

The authors are grateful to C.S.~Fischer and J.M.~Pawlowski for useful
discussions. H.G.~and J.J.~acknowledge financial support by the Deutsche
Forschungsgemeinschaft under contract Gi 328/1-2.

\begin{appendix}
\section{Non-Gau\ss ian gauge systems}

Up to this point, our analysis reveals that a construction of
standard-model-like theories based on non-Gau\ss ian $\hat\lambda$
fixed points and Gau\ss ian gauge fixed points still suffers from a
hierarchy problem in the four-fermion sector as well as triviality of
the abelian gauge sector -- only triviality in the Higgs-like sector
would be avoided in this scenario. In the following, we would like to
demonstrate that the existence of a non-Gau\ss ian fixed point in the
abelian gauge coupling has the potential to solve both problems
simultaneously.

Although the $\beta$ function for the abelian coupling $e^2$ as
derived from the mWTI does not furnish a non-Gau\ss ian fixed point
via the direct four-fermion contribution, such a fixed point might be
induced by the strong-coupling behavior of the gauge interactions or a
combination of strong gauge and four-fermion interactions. This would
manifest itself in a second zero of the anomalous dimension
$\etaB(e_\ast)=0$ at nonzero $e_\ast\neq0$. The search for such a
non-Gau\ss ian fixed point has a long tradition in the
literature. Lattice studies of non-compact abelian gauge systems using
staggered fermions \cite{Gockeler:1997dn} have not found such a fixed
point; on the contrary, numerical data is compatible with logarithmic
triviality. However, lattice results for field theories with both
gauge and four-fermion couplings are not yet available, although
perhaps accessible with recently developed algorithms
\cite{Kim:2001am}. Indications for the existence of such a fixed point
in a gauged NJL model have been collected using Dyson-Schwinger
equations \cite{Reenders:1999bg}. Of course, the UV behavior of system
as complex as the ones considered in this work is a completely open
problem.

Therefore, let us from now on assume that such a fixed point
$e_\ast>0$ in the abelian gauge coupling exists, and study its
consequences. Since the $\beta_{e^2}$ function for $e^2$ is positive
for small coupling, this non-Gau\ss ian fixed point is necessarily UV
stable, which solves the triviality problem. In order to study the
hierarchy problem, we have to compute the dependence of the critical
exponents for the $\hat\lambda$'s on $e_\ast$. For a first impression,
we perform the same numerical analysis as in
Sect.~\ref{sec::vanishing}, but now with nonzero $e=e_\ast$. Thereby,
we neglect any possible influence of the contributions of the unknown
$\beta_{e^2}$ function on the off-diagonal elements of the stability
matrix; this is justified if the $\hat\lambda$ couplings do not
play a dominant role for $\beta_{e^2}$ near $e=e_\ast$.

\begin{figure}[t]
\begin{center}
\subfigure{ \scalebox{0.6}[0.6]{
\begin{picture}(320,145)
\Text(261,30)[c]{\scalebox{1.4}[1.4]{$\alpha_{\ast}$}}
\Text(55,158)[c]{\scalebox{1.4}[1.4]{$\Theta_{\text{max}}$}}
\includegraphics{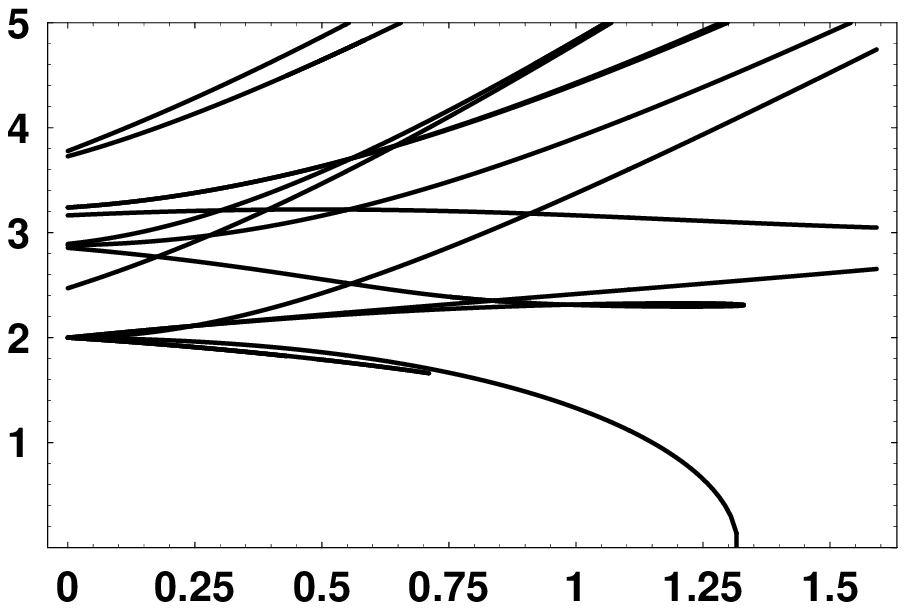}
\end{picture}}\label{unterm}}
\subfigure{ \scalebox{0.6}[0.6]{
\begin{picture}(320,145)
\Text(261,30)[c]{\scalebox{1.4}[1.4]{$\alpha_{\ast}$}}
\Text(55,158)[c]{\scalebox{1.4}[1.4]{$\Theta_{\text{max}}$}}
\includegraphics{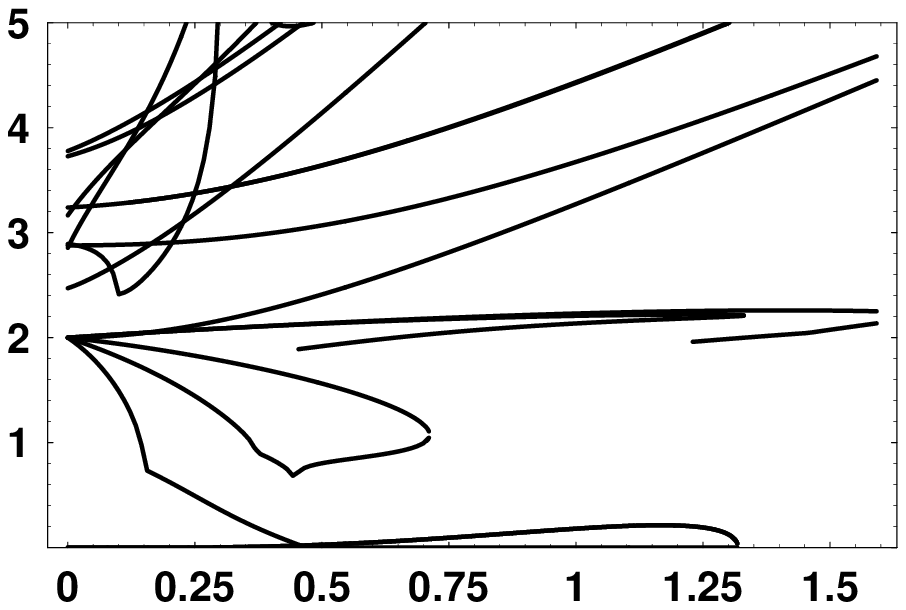}
\end{picture}}
\label{miteg}}
\end{center}

\vspace{-.5cm}
\caption{Maximal eigenvalue $\Theta_{\text{max}}$ of the various fixed
points of the four-fermion coupling depending on the assumed
fixed-point value $\alpha_{e\ast}=\frac{e^{2}_\ast}{4\pi}$ for the
abelian gauge coupling (at
$\alpha_{g\ast}=\frac{g^{2}_\ast}{4\pi}=0$). The left panel shows the
eigenvalue of the submatrix in the pure four-fermion sector while the
right panel depicts the eigenvalue for the full matrix but with the
unknown matrix elements in the pure gauge sector equal to zero.}
\label{fighierarchy}
\end{figure}

In Fig.~\ref{fighierarchy}, we display the dependence of the critical
exponents on the hypothetical value of the fixed point $e_\ast$. Each
line in these plots represents the maximal (most unstable) eigenvalue
for a given non-Gau\ss ian fixed point. As can be seen from the left
panel \ref{unterm}, there is one non-Gau\ss ian fixed point whose
maximal eigenvalue decreases with increasing $e_\ast$ and can become
close to zero if $\alpha_\ast\lesssim\alpha_{\text{cr}}\simeq1.3$. At
$\alpha_{\text{cr}}$, this fixed point annihilates with the Gau\ss ian
fixed point and disappears from the physically acceptable set.

In general, Fig.~\ref{fighierarchy} does not represent the complete
situation, since four-fermion couplings and gauge couplings mix
nontrivially at the non-Gau\ss ian fixed points. Technically speaking,
we should not neglect the $\partial (\pat e^2)/\partial \lambda$
contributions to the stability matrix, nor $\partial (\pat
(e^2,g^2))/\partial (e^2,g^2)$. Whereas we can read off the former
from the mWTI\re{betaeq}, nothing is known about the latter near the
speculative fixed point $e_\ast$. Since we do not want to introduce a
fine-tuning problem through the backdoor in this sector, it is natural
to assume that these entries in the stability matrix are small. If
this assumption is not valid, a large maximal eigenvalue will
probably arise from this sector, and the present speculation is
meaningless.

Therefore, we simply set the pure gauge entries to zero and study the
evolution of the eigenvalues including the $\partial (\pat
e^2)/\partial \lambda$ terms.  The result is shown on the right panel
of Fig.~\ref{fighierarchy}. Obviously, the mixing between the
couplings exerts a strong quantitative influence on the
eigenvalues. We find a whole range of possible $e_\ast$ fixed-point
values for which the maximal eigenvalue of the stability matrix is
small. The existence of such a non-Gau\ss ian gauge fixed point
therefore has the potential to stabilize the flow towards the IR
significantly. The running towards the Fermi scale would then proceed
with a small power or even almost logarithmically as for a system with
marginal couplings only. In such a scenario, the hierarchy problem
would be absent.

\end{appendix}

\end{document}